\documentclass[conference]{IEEEtran}
\IEEEoverridecommandlockouts
\usepackage{cite}
\usepackage{amsmath,amssymb,amsfonts}
\usepackage{algorithmic}
\usepackage{graphicx}
\usepackage{textcomp}
\usepackage{xcolor}
\def\BibTeX{{\rm B\kern-.05em{\sc i\kern-.025em b}\kern-.08em
    T\kern-.1667em\lower.7ex\hbox{E}\kern-.125emX}}

\usepackage{url}

\usepackage[english]{babel}
\usepackage{csquotes}

\usepackage{listings}
\usepackage{xcolor}

\newcommand{\RNum}[1]{\uppercase\expandafter{\romannumeral #1\relax}}

\newcommand{\tabitem}{~~\llap{\textbullet}~~}

\definecolor{codegreen}{rgb}{0,0.6,0}
\definecolor{codegray}{rgb}{0.5,0.5,0.5}
\definecolor{codeblue}{rgb}{0.016,0.52,0.86}
\definecolor{codered}{rgb}{0.96,0.047,0.24}
\definecolor{backcolour}{rgb}{0.97,0.97,0.97}

\lstdefinestyle{mystyle}{
  backgroundcolor=\color{backcolour},   commentstyle=\color{codegreen},
  keywordstyle=\color{codeblue},
  numberstyle=\tiny\color{codegray},
  stringstyle=\color{codered},
  basicstyle=\ttfamily\footnotesize,
  breakatwhitespace=false,         
  breaklines=true,                 
  captionpos=b,                    
  keepspaces=true,                 
  numbers=left,                    
  numbersep=5pt,                  
  showspaces=false,                
  showstringspaces=false,
  showtabs=false,                  
  tabsize=2
}
\usepackage[acronym]{glossaries}
\usepackage[acronym]{glossaries}

\newacronym{DMCs}{DMCs}{Data Matrix Codes}
\newacronym{QR-Codes}{QR-Codes}{Quick Response Codes}
\newacronym{IPC}{IPC}{Personal Computer}
\newacronym{IaC}{IaC}{Infrastructure as Code}
\newacronym{EAN}{EAN}{European Article Numbers}
\newacronym{UPC}{UPC}{Universal Product Codes}
\newacronym{AI}{AI}{Artificial Intelligence}
\newacronym{R-CNN}{R-CNN}{Region Based Convolutional Neural Network}
\newacronym{CNN}{CNN}{Convolutional Neural Network}
\newacronym{SSD}{SSD}{Single Shot Detector}
\newacronym{YOLO}{YOLO}{You Only Look Once}
\newacronym{IoU}{IoU}{Intersection over Union}
\newacronym{NMS}{NMS}{Non Maximum Suppression}
\newacronym{mAP}{mAP}{Mean Average Precision}
\newacronym{AP}{AP}{Average Precision}
\newacronym{OBB}{OBB}{Oriented Bounding Boxes}
\newacronym{GIoU}{GIoU}{Generalized Intersection over Union}
\newacronym{E-ELAN}{E-ELAN}{Extended Efficient Layer Aggregation Network}
\newacronym{BoF}{BoF}{Bag of Freebies}
\newacronym{GPU}{GPU}{Graphics Processing Unit}
\newacronym{CUDA}{CUDA}{Compute Unified Device Architecture}
\newacronym{Cae}{Cae}{Convolutional Autoencoder}
\newacronym{MLP}{MLP}{Multi Layer Perzeptron}
\newacronym{DIBCO}{DIBCO}{Document Image Binarization Contest}
\newacronym{RLD-Bild}{RLD-Bild}{Relative Darkness Feature-Bild}
\newacronym{GANs}{GANs}{Generative Adverserial
Networks}
\newacronym{P-FM}{P-FM}{Pseudo F-Measure}
\newacronym{FM}{FM}{F-Measure}
\newacronym{PSNR}{PSNR}{Peak Signal to Noise Ratio}
\newacronym{DRDM}{DRDM}{Distance-reciprocal Distortion Measure}
\newacronym{MSE}{MSE}{Mean-Squared-Error}
\newacronym{BCE}{BCE}{Binary-Cross-Entropy}
\newacronym{BCEWithLogits}{BCEWithLogits}{Binary-Cross-Entropy with Logits}
\newacronym{NdC}{NdC}{Number of decoded codes}
\newacronym{API}{API}{Application
Programming Interface}
\newacronym{AWS}{AWS}{Amazon Web Services}
\newacronym{EC2}{EC2}{Elastic Computing}
\newacronym{EKS}{EKS}{Elastic Kubernetes Service}
\newacronym{ECS}{ECS}{Elastic Container Service}
\newacronym{ECR}{ECR}{Elastic Container Registry}
\newacronym{IAM}{IAM}{Identity and Access Management}
\newacronym{RAM}{RAM}{Random Access Memory}
\newacronym{HTTP}{HTTP}{Hypertext Transfer Protocol}
\newacronym{VM}{VM}{Virtual machine}
\newacronym{YAML}{YAML}{Yet Another Markup
Language}
\newacronym{JSON}{JSON}{JavaScript Object Notation}
\newacronym{HCL}{HCL}{HashiCorp Configuration Language}
\newacronym{NAT}{NAT}{Network Address Translation}
\newacronym{HPA}{HPA}{Horizontal Pod Autoscaler}
\newacronym{OIDC}{OIDC}{OpenID Connect}
\newacronym{CLI}{CLI}{Command Line Interface}

\usepackage{pgfplots}
\pgfplotsset{compat=1.18}

\usepackage{multicol}

\begin{document}

\title{Comparison of Autoscaling Frameworks for Containerised Machine-Learning-Applications in a Local and Cloud Environment}

\author{\IEEEauthorblockN{1\textsuperscript{st} Christian Schröder}
\IEEEauthorblockA{\textit{Manufacturing Technologies} \\
\textit{Vitesco Technologies GmbH}\\
Limbach-Oberfrohna, Germany \\
cschroeder.research@gmail.com}
\and
\IEEEauthorblockN{2\textsuperscript{nd} René Böhm}
\IEEEauthorblockA{\textit{Manufacturing Technologies} \\
\textit{Vitesco Technologies GmbH}\\
Limbach-Oberfrohna, Germany \\
rene.boehm@vitesco.com}
\and
\IEEEauthorblockN{3\textsuperscript{rd} Alexander Lampe}
\IEEEauthorblockA{\textit{Dept. Engineering} \\
\textit{University of Applied Sciences}\\
 Mittweida, Germany \\
lampe@hs-mittweida.de}
}

\maketitle

\begin{abstract}
When deploying machine learning (ML) applications, the automated allocation of computing resources - commonly referred to as autoscaling - is crucial for maintaining a consistent inference time under fluctuating workloads. The objective is to maximize the Quality of Service metrics, emphasizing performance and availability, while minimizing resource costs. In this paper, we compare scalable deployment techniques across three levels of scaling: at the application level (TorchServe, RayServe) and the container level (K3s) in a local environment (production server), as well as at the container and machine levels in a cloud environment (Amazon Web Services Elastic Container Service and Elastic Kubernetes Service). The comparison is conducted through the study of mean and standard deviation of inference time in a multi-client scenario, along with upscaling response times. Based on this analysis, we propose a deployment strategy for both local and cloud-based environments.
\end{abstract}


\section{Introduction}
ML models can be utilized for anomaly detection in automated optical quality controls or time
series data analysis for manufacturing processes such as screwing, welding, or mounting. The
deployment of these models must adhere to Quality of Service requirements, encompassing
performance metrics like inference times and availability metrics related to upscaling
response times. Additional challenges \cite{ChallengesDeployment} arise during the transition of software from the development environment to the production environment. The IT department's support for software solutions in a production setting necessitates standardized, production-tested frameworks. With 1,416 logos featured on the 2023 Machine Learning, Artificial Intelligence, and Data Landscape \cite{MadLandscape}, establishing a standard proves to be a significant challenge.\\
Autoscaling has become a highly researched field, with a predominant focus on the development of autoscaling rules. Common techniques involve the use of machine learning algorithms such as reinforcement learning \cite{Khaleq}\cite{Hamid}\cite{Khaleq2}\cite{Joyce}, decision trees \cite{Abdullah}\cite{Vu} or long short term models \cite{Marie-Magdelaine}. Other methods are based on multi-level metric monitoring \cite{Taherizadeh}, fuzzy logic \cite{Tseng}, second order autoregressive moving average (ARMA) \cite{Roy} and simple moving average (SMA) \cite{Cheng}. Additionally, combinations of local predictors, including support vector machines and ARMA \cite{Kim} or long short term memory neural networks and autoregressive integrated moving average \cite{Radhika} are employed. While most autoscaling techniques focus on horizontal scaling, vertical scaling \cite{Jang} and combinations of both \cite{Kwan}\cite{Ding} are also of great interest. Reference \cite{Podolskiy} compares the autoscaling services of three cloud provider (Amazon Web Services, Microsoft Azure, Google Cloud Platform) on virtual machine (VM) and container level. To our best knowledge, there has been no research conducted to compare container and application-level autoscaling in an ML-specific task in both local and cloud environments.\\
This paper focuses on the use of well-known server and container orchestration frameworks. These frameworks include ML-specific webservers equipped with autoscaling capabilities, such as TorchServe and RayServe. Additionally, scalable container frameworks like Kubernetes and cloud services such as Amazon Web Services (AWS) Elastic Kubernetes Service (EKS) and AWS Elastic Container Service (ECS) are investigated. This paper does not consider cloud service-based, fully managed frameworks, such as AWS Sagemaker.\\
The contribution of this paper is to
\begin{itemize}
    \item Compare autoscaling-capable deployment methods using production-suitable frameworks on different scaling levels by deploying a multi-model application with conditional pre- and post-processing.
    \item Propose an optimal deployment method given the investigated frameworks for both local and cloud environments.
    \item Analyze potential performance improvements of RayServe concerning resource usage and suitability for simple model deployment.
\end{itemize}
The remaining sections of this paper are structured as follows. Section two introduces the real-world scenario of a multi-model application and describes the deployment methods investigated on three different levels of scaling. Section three compares the selected methods in a multi-client scenario. The fourth section offers further analysis of potential performance improvements of RayServe and a comparison of ML-specific frameworks for the use case of a simple model deployment. Finally, section five provides a summary of the results and outlines the next steps.
\section{Deployment Methods}
\begin{figure}[b]
    \centering
    \includegraphics[scale=0.45]{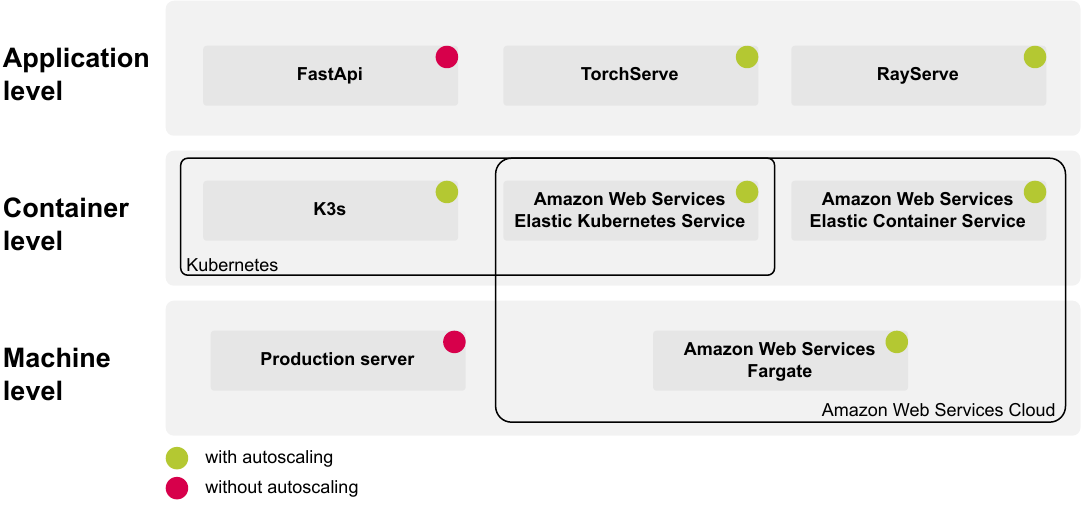}
    \caption{Frameworks and scaling level}
    \label{fig:ScalingLevel}
\end{figure}

The context involves the deployment of a machine learning (ML)-based multi-model application designed for decoding barcodes, data-matrix codes, and QR codes, with the combined size of all dependencies amounting to 1 GB. The process comprises three stages:
\begin{enumerate}
    \item Object Detection: Utilizing YOLOv5 to detect the code in the image.
    \item Code-Type Dependent Image Preprocessing: Involving rule-based image processing and custom ML models.
    \item Retrieving the Coded Information: Using an open-source decoding library.
\end{enumerate}
All machine learning models are implemented using PyTorch. The application's functionality is exposed through an Hypertext Transfer Protocol (HTTP) \acrfull{API}. To ensure consistent performance over time, irrespective of the request load, it is essential to scale the number of instances of the tool. Scaling can be achieved on three different levels, as illustrated in Fig. \ref{fig:ScalingLevel}, using the respective frameworks. FastApi is employed for all container-level scaling investigations, and multi-level scaling is not directly considered.
\subsection{Scaling on Application Level}
RayServe and TorchServe are deployment frameworks tailored for ML models. They offer automated horizontal scaling by incorporating new software units, ensuring that parallel requests are automatically distributed to different instances. However, the prerequisite for this is the availability of the required computing resources.\\
RayServe is framework-agnostic, emphasizing the ability to compose models, making it well-suited for the present use case. On the other hand, TorchServe specializes in ML models built with PyTorch, providing model optimization and native batch inference integration. With the use of custom model handlers, TorchServe can also be employed for model composition and complex pre- and post-processing.\\
An equivalent cloud service is represented by AWS Lambda. However, it cannot be used in this context due to its restriction on a maximum layer size of 250 MB \cite{AwsLambda}.
\begin{table*}[t]
\caption{Deployment components on Container level}
\label{tab:ContainerLevelDeploymentComponents}
  \centering
    \begin{tabular}{|l|l|l|}
    \hline
    \multicolumn{2}{|c|}{Components} &Function\\
    AWS ECS &Kubernetes (K3s, EKS) &\\
    \hline
     Cluster &Namespace &Logical grouping of services \\
     Service &Service &Grouping of Tasks/Pods of the same type under one exposed network address \\
     Task &Deployment/Pod &The running instance of one container \\
     Task Definition &Deployment Configuration &Description of the container runtime including:\\
      & &\tabitem Container image to use\\
      & &\tabitem Minimum and maximum CPU and memory resources\\
      & &\tabitem Networking mode to use\\
      & &\tabitem The startup command of the container\\
      & &\tabitem Environment variables\\
     Application Load Balancer & Ingress Controller &URL-based \acrfull{HTTP} routing from the outside client to\\
     & &the Kubernetes service\\
     Service Autoscaling &Horizontal Pod Autoscaler &Automatic scaling of workload components (Task, Deployment) based on network \\
      & &traffic or resource utilization\\
      Listener &Ingress Controller Specification &Defines rules for incoming requests (protocol, ip-adresses, ports) and routing\\
      Target Group &- &Routes requests to registered targets (e.g. AWS ECS container instances)\\
       & &and checks their health\\
      AWS CloudWatch &Metrics Server &Server for collecting and provisioning resource metrics (e.g. cpu, memory utilization)\\
    \hline
    \end{tabular}
\end{table*}
\subsection{Scaling on Container Level}
\begin{figure}[b]
    \centering
    \includegraphics[scale=0.34]{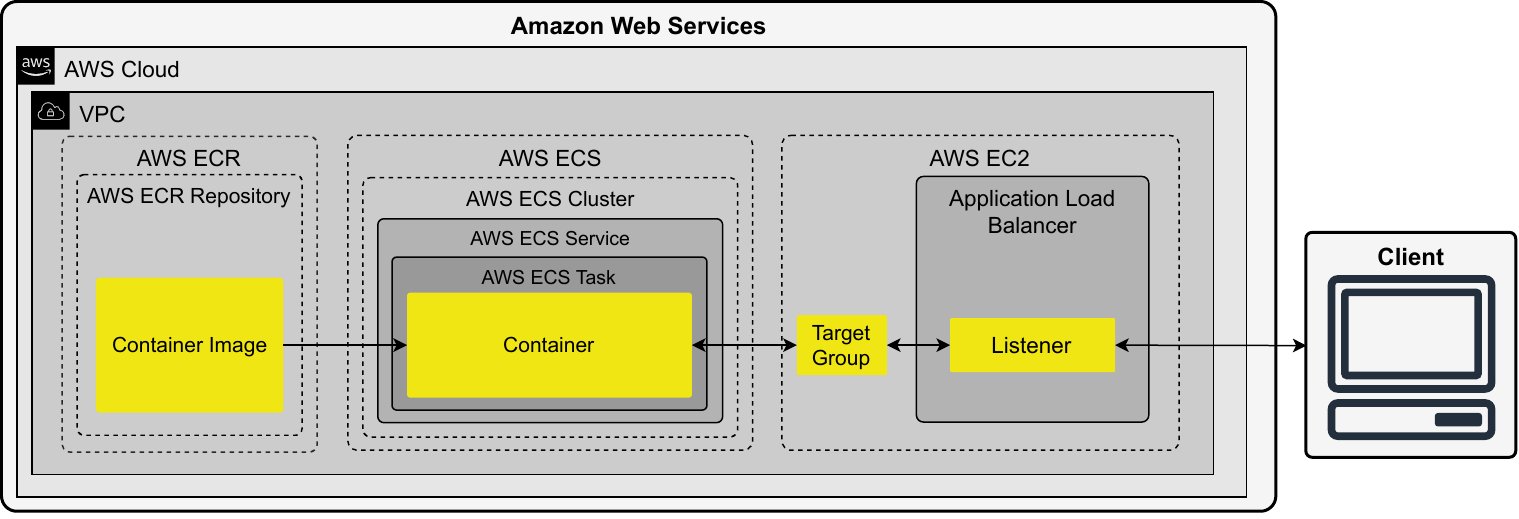}
    \caption{AWS ECS architecture}
    \label{fig:AwsEcsBackend}
\end{figure}

To facilitate container-level scaling, a Docker image is constructed with FastApi as the running web framework. Deploying software and its dependencies in a Docker container establishes a closed unit, allowing the application to run flexibly and hardware-agnostically. Containerization, using virtualization at the operating system level, enables scaling without replicating the entire operating system. A container image includes program code, dependencies (libraries), and runtime configuration parameters.\\
Kubernetes serves as a container orchestration tool, compatible with an open-source distribution like K3s or as part of a cloud service such as AWS EKS. AWS ECS is a cloud service with similar functionality, and both are utilized in this example with the underlying serverless compute engine AWS Fargate.\\
Fig. \ref{fig:AwsEcsBackend} illustrates the components needed to deploy a containerized application using AWS ECS. The container to be deployed is stored in a container repository provided by the AWS Elastic Container Registry (ECR) cloud service. The AWS ECS, the actual cloud deployment service, encompasses a cluster that logically groups services. Services group tasks under a single exposed network address. The ECS service manages scaling of container instances through its autoscaling function. To access the service externally, an application load balancer (ALB) is essential. As part of the AWS EC2 service, the ALB routes requests to one or more healthy target groups. The listener, as part of the ALB, defines rules for the handling and routing of incoming requests. A target group acts as an additional distributor, accepting requests from the ALB and forwarding them to registered and healthy targets, such as the ECS service. The architecture of Kubernetes is analogous to that of AWS ECS, with a comparison of elements provided in Table \ref{tab:ContainerLevelDeploymentComponents}.

\subsection{Scaling on Machine-Level}
Virtual machine scaling occurs automatically when scaling at the container level in the cloud. AWS ECS and EKS can be used serverless with AWS Fargate. Since this service is fully managed by AWS, no further analysis can be conducted on this topic.
\section{Comparison of selected deployment methods}
The test scenario involves four clients simultaneously sending 1000 requests to the deployed decoding application described earlier.\\
The scaling behavior of AWS ECS, AWS EKS and K3s is configured to be based on average CPU utilization with a threshold of 50\%. The scaling behaviour of RayServe is configured based on the target number of ongoing requests being one. However, TorchServe cannot be configured in a similar way. For all frameworks, a minimum of one instance and a maximum of four instances are configured.\\
To prevent interference with the autoscaling frameworks, parallelism throughout PyTorch must be prohibited. This is achieved by setting the number of threads used by PyTorch to one.\\
AWS Fargate serves as the execution environment for the AWS ECS and EKS containers. The K3s, RayServe, and TorchServe frameworks are executed on an AWS EC2 virtual machine instance of type m4.4xlarge with 16 virtual CPUs (vCPU). Requests are served from four clients running on the same virtual machine to minimize network latency.\\
When using AWS ECS and EKS along with Fargate, each container instance is assigned one CPU core and two GB of RAM. This is achieved for AWS ECS by setting the appropriate values in the task definition. The Kubernetes distributions (EKS, K3s) limit the resource usage of their pods by the same measure in the corresponding deployment configuration file. The minimum CPU requirements are set at 900m (meaning 900 millicpu or 90 \% of the execution time of a CPU core) and 1024Mi (meaning 1024 mebibytes or 1074 megabytes) for memory. The maximum limits are set at 1000m for CPU and 2048Mi for memory.\\
Due to the restriction that both (RayServe and TorchServe) application instances cannot be limited independently, their corresponding containers are limited to a resource usage of $k=5$ CPU cores and ten GB of memory. The choice of five CPU cores is made to provide one CPU core for each desired instance of the application and one dedicated CPU core for the scaling framework.\\

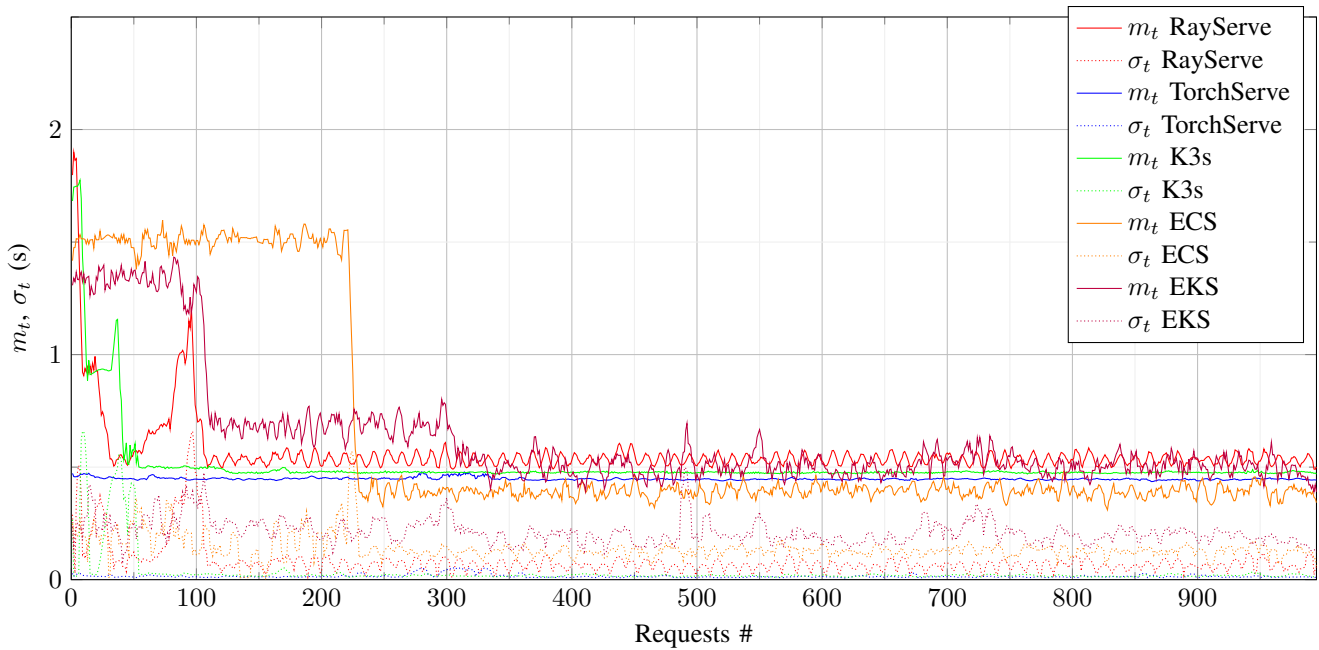
\begin{figure*}[t]
\centering
\pgfplotstableread{data2.dat}{\mydataz}
\begin{tikzpicture}
\begin{axis}[
    legend style={at={(0.99,0.72)},anchor=east},
    xmin = 0, xmax = 995,
    ymin = 0, ymax = 2.5,
    xtick distance = 100,
    ytick distance = 1,
    grid = both,
    minor tick num = 1,
    major grid style = {lightgray},
    minor grid style = {lightgray!25},
    width = \textwidth,
    height = 0.5\textwidth,
    xlabel = {Requests \#},
    ylabel = {$m_t$, $\sigma_t$ (s)},
    legend cell align={left},]

% Plot data from a file
\addplot[red] table [x = {x}, y = {y1}] {\mydataz};
\addplot[red, densely dotted] table [x ={x}, y = {y2}] {\mydataz};
\addplot[blue] table [x = {x}, y = {y3}] {\mydataz};
\addplot[blue, densely dotted] table [x ={x}, y = {y4}] {\mydataz};
\addplot[green] table [x = {x}, y = {y5}] {\mydataz};
\addplot[green, densely dotted] table [x ={x}, y = {y6}] {\mydataz};
\addplot[orange] table [x = {x}, y = {y7}] {\mydataz};
\addplot[orange, densely dotted] table [x ={x}, y = {y8}] {\mydataz};
\addplot[purple] table [x = {x}, y = {y9}] {\mydataz};
\addplot[purple, densely dotted] table [x ={x}, y = {y10}] {\mydataz};
\legend{
    $m_t$ RayServe,
    $\sigma_t$ RayServe,
    $m_t$ TorchServe,
    $\sigma_t$ TorchServe,
    $m_t$ K3s,
    $\sigma_t$ K3s,
    $m_t$ ECS,
    $\sigma_t$ ECS,
    $m_t$ EKS,
    $\sigma_t$ EKS,
}
\end{axis}
\end{tikzpicture}
\caption{Inference times at scaling test for different deployment methods}
\label{fig:Inferenzzeiten Skalierung}
\end{figure*}
The comparison of the scaling behaviour of the five methods studied:
\begin{itemize}
    \item Scaling on application level with 
    \begin{itemize}
        \item RayServe
        \item TorchServe
    \end{itemize}
    \item Scaling on container level with
    \begin{itemize}
        \item K3s
        \item AWS ECS
        \item AWS EKS
    \end{itemize}
\end{itemize}
is based on the moving mean $m_t$ and standard deviation $\sigma_t$ of the inference time for five consecutive requests from each client, as illustrated in the chart in Fig. \ref{fig:Inferenzzeiten Skalierung}. Key observations include:
\begin{itemize}
    \item The local scaling methods (RayServe, TorchServe, K3s) adapt most rapidly to the request load since they already possess the necessary additional computing resources.
    \item Cloud services (AWS ECS, AWS EKS) exhibit slower scaling as they involve machine-level scaling in the background, requiring the initiation of new virtual machines. There is also a delay due to the granularity of AWS CloudWatch metrics, set at one minute for metrics produced by AWS.
    \item AWS EKS initiates scaling earlier than AWS ECS. However, AWS EKS scales from one to three instances and later to four instances, while AWS ECS scales directly to four instances.
    \item Both Kubernetes distributions (AWS EKS, K3s) exhibit comparable incremental scaling behavior, increasing from one to two and later to four instances (K3s) or from one to three and then to four instances (EKS).
\end{itemize}
\begin{table}[b]
\caption{Inference times in a scaled condition (multi-model application)}
\label{tab:Inference Times scaled condition}
  \centering
    \begin{tabular}{|c|c|c|c|c|c|}
    \hline
    &RayServe &TorchServe &K3s &ECS &EKS\\
    \hline
    $m_{t}$ [s] &0.534	&0.445	&0.476	&\textbf{0.393}	&0.503\\
    $\sigma_t$ [s] &0.063	&\textbf{0.012}	&0.017	&0.126	&0.198\\
    \hline
    \end{tabular}
\end{table}
An additional analysis is conducted using the mean and standard deviation of the inference times in a scaled condition starting at request 400 for all methods. The results in Table \ref{tab:Inference Times scaled condition} reveal the following insights:
\begin{itemize}
    \item Regarding cloud services, ECS demonstrates a significantly smaller mean and standard deviation of inference times compared to EKS.
    \item Regarding local frameworks, TorchServe exhibits the smallest mean and standard deviation of inference times, taking 17 \% less time than RayServe.
    \item All local frameworks demonstrate a lower standard deviation of inference times than the cloud services.
\end{itemize}
Additional notes:
\begin{itemize}
    \item Comparing local and cloud services, no significant difference in network speed is expected, given that the client is in the same cloud region as the deployment service.
    \item Assuming identical computing time and network speed based on identical computing hardware, differences in inference time for all locally executed frameworks are attributed to the handling of the frameworks' main tasks, including \cite{AutoscalingKubernetes}:
    \begin{itemize}
        \item Monitoring workloads
        \item Analyzing the monitored data
        \item Deciding on suitable scaling methods
        \item Allocating resources 
        \item Starting up or shutting down instances
        \item Distributing the requests to the several instances
    \end{itemize}
\end{itemize}
In conclusion, in the context of a multi-model application with a size of 1 GB, both TorchServe and K3s are equally recommended for local deployment. In cloud environments, AWS ECS is favored over AWS EKS due to superior performance. 

\section{In depth analysis in local environment}
RayServe is a frequently employed framework in the industry, attributed to its framework-agnostic design and seamless integration of advanced HTTP methods via FastApi. Consequently, an additional evaluation is undertaken:
\begin{enumerate}
	\item To scrutinize potential enhancements in performance concerning resource utilization.
	\item To assess its competitive performance in the prevalent use case of deploying a single ML model.
\end{enumerate}
\subsection{RayServe resource usage}
The scaling behavior of RayServe, concerning the multi-model application, is scrutinized in terms of resource allocation impact on the inference times of concurrently processed requests. The assessment is conducted on an AWS EC2 virtual machine of type m4.4xlarge equipped with 16 vCPUs. Parallel client processes, denoted by $k$ and varying within the range $k\in[1, 2, 4]$, are established. Each client process sends 100 requests to the pre-scaled RayServe server, operating within a Docker container with $n$ CPU cores allocated, where $n$ spans the range $n\in[k, k+1, k+2, k+3]$. 
Table \ref{tab:Inferenzzeiten RayServe CPU} outlines the results, revealing the following observations:
\begin{itemize}
    \item Utilizing $n=k+1$ CPU cores manifests substantial improvements in inference time compared to $n=k$ CPU cores.
    \item With more than $n=k+1$ CPU cores, a saturation point is reached, leading to diminishing returns in the enhancement of inference time.
    \item Parallel requests consistently exhibit significantly higher inference times than the sequential processing of requests from a single client.
\end{itemize}
From these findings, the following conclusions can be drawn:
\begin{itemize}
    \item The RayServe environment attains optimal performance when allocated at least $n = k+1$ CPU cores for $k$ clients concurrently operating.
    \item The cost-optimal configuration is precisely $n = k+1$ CPU cores, ensuring a dedicated CPU core for the RayServe controller.
    \item RayServe demonstrates an inability to deliver consistent inference times irrespective of the number of clients, even with additional hardware resources.
\end{itemize}
\begin{table}[t]
\caption{RayServe inference times depending on the number of CPU cores}
\label{tab:Inferenzzeiten RayServe CPU}
  \centering
    \begin{tabular}{|c|c|c|c|c|}
    \hline
    k &n=k &n=k+1 &n=k+2 &n=k+3\\
    \hline
    1 &0.636 s &0.478 s &0.467 s &0.479 s\\
    2 &0.606 s &0.490 s &0.485 s &0.490 s\\
    4 &0.608 s &0.530 s &0.509 s &0.504 s\\
    \hline
    \end{tabular}
\end{table}

\subsection{Simple model deployment}
While the capability to compose models exists within ML-specific frameworks, the deployment of model compositions can be seamlessly achieved using a generic webserver, subsequently scaled in containers, as previously demonstrated. To delve into a more straightforward use case, the deployment of a single ML model is investigated. In this context, an EfficientNet-B0 model is trained for classification utilizing Pytorch. The neural network is serialized as a TorchScript model. The comparison encompasses the following frameworks and configurations:
\begin{itemize}
    \item RayServe
    \item TorchServe
    \item TorchServe with Batch Inference
    \item FastApi
\end{itemize}
Each framework employs identical source code for model loading, as well as pre- and post-processing. Execution within a Docker container is standardized to ensure uniform resource usage, specifically limiting it to $n = 5$ CPU cores. For the ML-specific frameworks, a scaled condition is established from the outset, with four replicas/workers set as both the minimum and maximum. Owing to the absence of autoscaling considerations, the underlying web framework of RayServe, FastApi, is leveraged for comparison. The application server, Gunicorn, employed in the FastAPI framework, is configured to operate with a worker count determined by the formula (2 x num cores) + 1 = 11, in accordance with the recommendations provided in the documentation \cite{GunicornWorkers}.\\
The test scenario involves four clients concurrently dispatching 100 requests each, running on the same virtual machine as the Docker container. The batch inference feature in TorchServe utilizes a batch size of four with a maximum batch delay of 40 milliseconds, determining the time to wait until the defined batch size is attained.\\ 
The mean and standard deviation of inference times are presented in Table \ref{tab:Inference Times scaled condition Simple Deployment}, revealing the following observations:
\begin{itemize}
    \item RayServe exhibits the highest mean and standard deviation among all frameworks, requiring on average 16 \% and 42 \% more time than FastApi/TorchServe, respectively.
    \item TorchServe with batch inference demonstrates significantly faster performance than TorchServe in its default configuration
    \item TorchServe with batch inference exhibits a threefold increase in the standard deviation of inference times compared to its default configuration.
\end{itemize}
RayServe manifests notably higher inference times when deploying a single model, contrasting with TorchServe's superior performance, being consistent with the observations from the examination of multi-model application deployment in Fig. \ref{fig:Inferenzzeiten Skalierung}. TorchServe exhibits superior performance again, surpassing even the high-performance \cite{FastApi} web framework FastApi. The incorporation of batch inference further elevates its efficiency. In the current testing scenario with numerous clients and a consistent request load, the mean inference time experiences a 23 \% reduction. It is imperative to note that scenarios with less constant request loads or fewer clients than the batch size may lead to an increase in inference time due to batch delay. This variability is also reflected in the augmented standard deviation of batch inference times.
\begin{table}[t]
\caption{Comparison of inference times in a scaled condition (simple model deployment)}
\label{tab:Inference Times scaled condition Simple Deployment}
  \centering
    \begin{tabular}{|c|c|c|c|c|c|c|}
    \hline
    metric &RayServe &TorchServe &TorchServe &FastApi\\
     & & &Batch Inference &\\
    \hline
    $m_{t}$ [s] 	&0.143	&0.101	&\textbf{0.068} &0.123\\
    $\sigma_t$ [s] &0.035	&\textbf{0.005}	&0.013 &0.012\\
    \hline
    \end{tabular}
\end{table}

\section{Conclusion and next steps}
The scalable deployment of a multi-model machine learning application, encompassing intricate image pre- and post-processing, prompts considerations regarding the optimal scaling level. Machine learning-specific frameworks, such as RayServe and TorchServe, offering autoscaling at the application level, have proven applicable for an application of the present size. TorchServe emerges as the frontrunner, showcasing the fastest inference times among all locally operating frameworks investigated. A parallel performance is observed with K3s, scaling locally at the container level. Consequently, the recommendation for local deployment encompasses the use of both TorchServe and FastApi with K3s.\\
In the AWS cloud environment, where AWS ECS and AWS EKS operate on the AWS Fargate serverless compute engine, comparable scale-up times are noted due to the necessity of initiating new underlying virtual machines. AWS ECS is favored over AWS EKS, primarily due to its significantly lower mean inference time.\\
An additional analysis evaluates the performance of RayServe in terms of resource utilization and the size of the deployed application. With k+1 CPU cores identified as the cost optimum for k clients, RayServe exhibits no further improvement with increased resources and lacks the capability to maintain a constant inference time independent of the number of clients. The use of a simple ML application reaffirms the earlier conclusion that TorchServe significantly outperforms RayServe.\\
Further investigations may include:
\begin{itemize}
    \item Partitioning a singular multi-model application into multiple scaling instances to augment scalability and enhance parallelizability.
    \item Implementing distributed/multi-node computing to conduct model inference on GPU instances and pre- and post-processing on cost-optimized CPU instances.
    \item Formulating autoscaling rules for the anticipation of the requisite number of instances in real-world scenarios.
    \item Integrating scaling mechanisms at both the application and container levels to ameliorate scaling dynamics and optimize costs.
\end{itemize}

\enlargethispage{-2.4in}
\bibliographystyle{IEEEtran}
\bibliography{literature}

\end{document}